\documentclass[12pt,reqno]{amsart}
\usepackage{amsfonts}
\usepackage[english]{babel} 
\usepackage{graphicx}
\usepackage{color}
\usepackage[utf8]{inputenc}
\newtheorem{theorem}{Theorem}[section]

\newtheorem{corollary}[theorem]{Corollary}

\newtheorem{remarks}[theorem]{Remark}

\usepackage{hyperref}

\begin{document}

%\begin{flushright}                                       
\begin{figure}[h]
%Centramos la figura
%\hspace{8cm}
%\begin{center}
% Incluimos el grafico y definimos el ancho y alto, la
% figura se encuentra en el fichero imagen.jpg
%\includegraphics[height=1cm,width=1.5cm]{MAT4.png}\\
Preprint DMATULS-01/2018-ULS-Chile\\
{\small Author@Corgini2018}\\
%{\small https://creativecommons.org/licenses/by-nc-nd/4.0/}\\
%\includegraphics[height=1.5cm,width=2cm]{lic.jpg} 
%\end{center}
\vspace{0cm}
\end{figure}
%\end{flushright}

\author{M. Corgini}
\title[]{NEW SCENARIO FOR THE EMERGENCE OF NON-CONVENTIONAL BOSE-EINSTEIN CONDENSATION. BEYOND THE NOTION OF  ENERGY GAP }

\address{Departamento de Matem\'aticas, Universidad de La Serena\\Cisternas 1200, La Serena. Chile. \url{http://www.dmatuls.cl/portal/}}  

\maketitle

\begin{abstract}
 A  stable non-ideal Bose system whose energy operator includes a  perturbations depending on the square root  of the number operator associa\-ted to the zero mode energy  is ana\-lyzed,  demonstrating  that, in pre\-sence or  absence of a gap in the one particle  energy spectrum,  it undergoes the so-called  non-conventional Bose–Einstein condensation.   
\end{abstract}

\bigskip

\section{Introduction}\label{sec:1}

\bigskip

\subsection{Grand canonical ensemble}

Let
\begin{equation*}
p_{\ell}(\beta, \mu) = \frac{1}{\beta V_{\ell}}\log \mathop{\rm Tr}\nolimits_{\mathcal{F%
}_B} \exp \left( -\beta( \hat{\Gamma}_{\ell}- \mu \hat{N})\right),
\end{equation*} be the grand-canonical pressure (grand canonical ensemble) at finite
volume related to a quantum many particle Bose system whose  Hamiltonian is represented by $\hat{\Gamma}_{\ell},$ being $\beta = \theta^{-1}$ 
the inverse temperature, $\mu$ the chemical potential, $\hat{N}$ the total number of particles operator and $$\mathcal{F}_B:=\displaystyle\oplus^{\infty}_{N=0} \mathcal{H}^{N}_B$$ the Fock space for bosons constructed from the Hilbert space $  \mathcal{H} = L^2(\Lambda_{\ell}),$ where $\Lambda_{\ell} \subset \mathbb{R}^3 $ is a cubic box of volume $V_{\ell}= {\ell}^3.$  In this case, $$\mathcal{H}^{N}_B := S \overbrace{\mathcal{H}\otimes\mathcal{H}...\otimes\mathcal{H}}^{N-{\rm times}}$$ represents the Hilbert space associated to a system composed by exac\-tly $N$ Bose-particles, being $S$  the symmetrization operator.

If the limit pressure $  p (\beta, \mu)= \displaystyle\lim_{V_{\ell}\to\infty}p_{\ell}(\beta, \mu)$ exists for 
all $\mu \in \mathbb{R}$ it  will be said that it is a superstable system. If this holds only for $\mu$ in a proper subset of $\mathbb{R},$ the system will be denominated ``stable".

Being $\hat{\Gamma}_{\ell}
(\mu)= \hat{\Gamma}_{\ell} - \mu \hat{N}$, the equilibrium Gibbs
state (grand canonical ensemble) $\langle -
\rangle_{\hat{\Gamma}_{\ell}
(\mu)}$ is defined as

\begin{equation*}
\langle \hat{A} \rangle_{\hat{\Gamma}_{\ell}(\mu)} = \left( \mathop{\rm Tr}\nolimits_{
\mathcal{F}_B} \exp \left( -\beta \hat{\Gamma}_{\ell} (\mu)\right)\right)^{-1}
\mathop{\rm Tr}\nolimits_{\mathcal{F}_B} \hat{A} \exp \left( -\beta \hat{\Gamma}
_{\ell}
 (\mu)\right),
\end{equation*}
for any operator $\hat{A}$ acting on $\mathcal{F}_B.$
The total density of particles $\rho(\mu)$ for infinite
volume is defined as
\begin{equation*}
\displaystyle\lim_{V_{\ell}\to\infty}\left\langle \frac{\hat{N}}{V_{\ell}}\right\rangle_{\hat{\Gamma}
_{\ell}
(\mu)} = \displaystyle\lim_{V_{\ell}
\to\infty}\rho_{\ell}(\mu) = \rho (\mu) =\mbox{constant}.
\end{equation*}

\bigskip

\subsection{Canonical ensemble} Let consider a sequence of finite systems enclosed in the cubic region $\Lambda_{\ell}.$ In the canonical ensemble framework, the free ener\-gy $f_{\ell}(\beta,\varrho_{\ell})$ associated to Hamiltonian $\hat{\Gamma}_{\ell}$  at finite number of particles $N,$  finite volume $V_{\ell}$, inverse temperature $\beta$ and density $\varrho_{\ell},$ is defined as: 

\[f_{\ell}(\beta,\varrho) = -\frac{1}{\beta V_{\ell}}\log \mathop{\rm Tr}\nolimits_{\mathcal{H}^{N}_B} \exp \left(-\beta \hat{\Gamma}^{(N)}_{\ell}\right),  \]where $\Gamma^{(N)}_{\ell}$ is the restriction of the energy operator to $\mathcal{H}^{N}_B,$ and 
\[ \varrho_{\ell} =\frac{ N}{V_{\ell}},\;\varrho =\displaystyle\lim_{V_{\ell}, N\to\infty}\varrho_{\ell} =\mbox{constant},\]

\[ \varrho_{{\mathbf{0}},\ell} ={n_{\mathbf{0}}}/{V_{\ell}},\; \varrho_ {\mathbf{0}}=\displaystyle\lim_{V_{\ell},N\to\infty} \varrho_{{\mathbf{0}},\ell},\] being 
$ n_{\mathbf{0}}$  the number of particles associated to the zero mode. 

In what follows, the symbols $\varrho,\varrho_{\mathbf{0}}$ will be used when referring
to $\varrho_{\ell}$ or $\varrho_{\mathbf{0},\ell}$ indistinctly, avoiding excessive
notation.

Finally, being $\hat{A}$ an arbitary operator defined on the Fock space $\mathcal{F}_B,$  the thermal average,  in the canonical ensemble,  of the operator $\hat{A}^{(N)} := \hat{A}|_{\mathcal{H}^{N}_B},$ at finite volume $V_{\ell}$, inverse temperature $\beta$ and density $\varrho,$ is defined as:   

\begin{equation*}
\langle \hat{A}^{(N)} \rangle_{\hat{\Gamma}^{(N)}_{\ell} (\varrho)}= \left( \mathop{\rm Tr}\nolimits_{
\mathcal{H}^{N}_B} \exp \left( -\beta \hat{\Gamma}^{(N)}_{\ell}\right)\right)^{-1}
\mathop{\rm Tr}\nolimits_{\mathcal{H}^{N}_B} \hat{A}^{(N)}  \exp \left( -\beta \hat{\Gamma}^{(N)}
_{\ell}\right).
\end{equation*} 

\bigskip

\subsection{Non-conventional Bose-Einstein condensation}

It will be said that a quantum many particle Bose system undergoes  non-conven\-tional Bose-Einstein condensation (NCBEC) if it  shows an independent  on temperature  macros\-copic occupation of
the single particle zero mode, i.e., $ \rho_{\mathbf{0}} (\mu) > 0.$  
  
On the other hand, the kinetic energy of a single particle in a  Bose system is described by the free Laplacian ope\-rator $\triangle_{\ell}$ defined on $\Lambda_{\ell}$  endowed with appropriate boundary conditions and  a numerable set of eigenvalues $E,$ obtained from the equation \[- \frac{1}{2}\triangle_{\ell} \phi = E\phi\] ($\phi$'s are the free Lapacian eigenfunctions). 

The non-conventional condensation can emerge, among other reasons, as consequence, either  from the existence in the thermodynamic limit  of an  isolated and negative eigenvalue (the smallest one) of the laplacian or by shifting the zero energy level of the kinetic energy ope\-rator to a negative value, generating a gap in its spectrum.

In the first case, the so-called attractive boundary conditions (see ref.~\cite{Rob1}), given by the condition: 

$$\frac{\partial\phi}{\partial n} + \sigma\phi = 0,$$ play a fundamental role.  $\frac{\partial\phi}{\partial n} $ is the outward normal derivative on the boundary and $\sigma< 0$ is the parameter governing the attractivity of the boundaries. In this context, the lower eigenvalue is an isolated negative point of the spectrum.

In the second case (periodic boundary conditions), the mentioned finite gap in the one-particle excitations spectrum can be obtained by adding to the corresponding energy operator -in the formalism of se\-cond quantization- the  term $\Delta\hat{n}_{\mathbf{0}},$ where $\Delta < 0.$ As an example,  a homogeneous Bose gas with periodic boundary conditions and a two-body interaction enclosed in a region of volume $V_{\ell}$ has been exhaustively studied  in refs.~\cite{LVZ1,LVZ2}. In these works the emergence of NCBEC  has been proven.

Unlike the described situations, the goal of this work is to demons\-trate that non-conventional macroscopic occupation of the fundamental state is possible, from a purely mathematical point of view, without resorting to the above mentioned arguments. Indeed,  in ref.~\cite{CT} it has been shown that a kind of small nonlinear perturbations of  the energy operator of the ideal Bose gas -represented by the square root  of the number operator associated to the zero mode energy-  in absence of any gap in the one particle  energy spectrum of the system, leads to the emergence of non-conventional Bose-Einstein condensation. This fact opens new scenarios for the ocurrence of this type of  critical behavior. 

This paper is organized as follows. Section~\ref{sec:2} is devoted to introduce  the  perturbed noninteracting Bose particle system to be stu\-died. In section~\ref{sec:3} the effects of the before mentioned disturbance on its critical behavior, in presence or absence of gaps, is  analized. Finally, sections  ~\ref{sec:4}  and ~\ref{sec:5} consider some remarks and the main conclusions, respectively.  

\bigskip

\section{Perturbed system}\label{sec:2}

\bigskip

\subsection{General framework}

Let consider a particle system enclosed within the cubic box $\Lambda_{\ell} \subset \mathbb{R}^3$  of volume $V_{\ell}.$  The free energy operator $\hat{H}^0_{\ell} $  is given as
\[\hat{H}^0_{\ell} = \displaystyle\sum_{\mathbf{p}\in {\Lambda
}^*_{\ell}}\lambda_{\ell}(\mathbf{p}) \hat{n}_{\mathbf{p}},\]  where, $ \hat{
N} = \displaystyle
\sum_{\mathbf{p}\in {\Lambda }^*_{\ell}} \hat{a}
^{\dag}_{\mathbf{p}} \hat{a}_{\mathbf{p}}$ is the total number operator. The sum  runs over the set $
\Lambda^*_{\ell} = \{ \mathbf{p}= (p_1,\dots,p_d)\in {\mathbb R}^d:
p_{\alpha}={2\pi n_{\alpha}}/{\ell}, n_{\alpha} \in {\mathbb Z}, \alpha
=1,2,\dots,d\}$. $\hat{a}^{\dag}_{\mathbf{p}},
\hat{a}_{\mathbf{p}}$ are the Bose ope\-rators of creation and
annihilation of particles defined on the Bose Fock space
${\mathcal{F}}_B,$  satisfying the usual commutation rules: $\lbrack \hat{a}
_{\mathbf{q}},\hat{a}^{\dag}_{\mathbf{p}} \rbrack
= \hat{a}_{\mathbf{q}}
\hat{a}^{\dag}_{\mathbf{p}} - \hat{a}^{\dag}_{\mathbf{p}} \hat{a}
_{\mathbf{q}} = \delta_{\mathbf{p},\mathbf{q}} I$ ($I$ is the identity operator),  and $\hat{n}
_{\mathbf{p}} = \hat{a}^{\dag}_{\mathbf{p}}
\hat{a}_{\mathbf{p}}$ is the number operator associated to
mode $\mathbf{p}.$ 
On the other hand, the operators $\hat{H}^{0 \prime}_{\ell}$ and $\hat{N}^{{\prime}}$   are defined as:

\bigskip

$$ \hat{H}^{0 \prime}_{\ell} =  \displaystyle\sum_{\mathbf{p}\in {\Lambda
}^*_{\ell}\backslash \{\mathbf{0}\}}
\lambda_{\ell}(\mathbf{p})\hat{n}_{\mathbf{p}},\;\;\hat{N}^{{\prime}} = \displaystyle\sum_{\mathbf{p}\in {\Lambda }
^*_{\ell}\backslash \{ \mathbf{0} \}
}\hat{a}^{\dag}_{\mathbf{p}}\hat{a}_{\mathbf{p}}.$$ 

\bigskip

Assuming  periodic boundary conditions, the energies $\lambda_{\ell}(\mathbf{p})$ asso\-ciated to the single modes are given by, 
\begin{equation*}
\lambda_{\ell}(\mathbf{p})= \left\{\begin{array}{ll}
\Delta < 0, & \mbox{ $ \mathbf{p}=\mathbf{0} $} \\
\noalign{\smallskip}{\mathbf{p}^2}/{2},&\mbox{ $\mathbf{p}\neq
\mathbf{0},$}
\end{array}\right. \end{equation*} where $\Delta$ has been arbitrarily introduced for producing a gap in the spectrum. 

\bigskip

\subsection{Problem statement}

Let $\hat{H}_{\ell}$ be  the following Hamiltonian:

\begin{equation}\begin{split}\label{eqn:1}  \hat{H}_{\ell} = \hat{H}^{0}_{\ell} + \frac{a}{V_{\ell}}\hat{a}^{\dag 2}_{{\mathbf 0}} \hat{a}^2_{{\mathbf 0}},\;\; a>0.
\end{split}\end{equation}

\bigskip

The second term on the right hand side of~(\ref{eqn:1}) can be physically
understood as the simultaneous creation of  2-single zero mode particles after
the disappearance of an equivalent amount of single particles associated to the
same mode. It is a well-known fact that the particle system with energy operator~(\ref{eqn:1}) -extensively studied in refs.~\cite{BZ1,BZ2,BZ3, BZ4,CS}  by using different mathematical techniques- undergoes NCBEC for $\mu  \in (\Delta, 0].$ 

This work deals with a modified version of the Hamiltonian given in~(\ref{eqn:1}), which consists  in including in it the disturbance $ -2\nu\sqrt{V_{\ell}\hat{n}_{\mathbf{0}}},$ being $ \nu > 0.$  In other words, the perturbed  energy operator has the following form: 

\begin{equation}\label{eqn:2}\begin{split} \hat{H}^{\rm per}_{\ell}  = \hat{H}^{0}_{\ell}  + \frac{a}{V_{\ell}}\hat{a}^{\dag 2}_{{\mathbf 0}} \hat{a}^2_{{\mathbf 0}}  - 2\nu\sqrt{V_{\ell}\hat{n}_{\mathbf{0}}}.
 \end{split}\end{equation}  

\bigskip

Since $ \hat{a}^{\dag 2}_{{\mathbf 0}} \hat{a}^2_{{\mathbf 0} }=\hat{n}_{\mathbf{0}}(\hat{n}_{\mathbf{0}} -I),$ $ \hat{H}^{\rm per}_{\ell}$ can be rewritten as:

\bigskip

\begin{equation}\label{eqn:3}\hat{H}^{\rm per}_{\ell} =\hat{H}^{0\prime}_{\ell}  +V_{\ell} \left(\Delta - \frac{a}{V_{\ell}}\right)\hat{n}_{\mathbf{0}} + \frac{a}{V_{\ell}}\hat{n}^2_{\mathbf{0}}  -2\nu\sqrt{V_{\ell}\hat{n}_{\mathbf{0}}}.\end{equation}

\bigskip

The main goal of this article  is to prove that such a perturbation leads  to the emergence of  non-conventional Bose–Einstein condensation in presence or absence of a gap in the kinetic energy spectrum.

\bigskip

\section{Emergence of non-conventional Bose-Einstein Condensation}\label{sec:3}

\bigskip

\begin{theorem}\label{thm:1} Let  $ \mu \leq 0,\;\nu > 0.$ Then,  the limit pressure $p^{\rm per} (\beta,\mu, \nu, \Delta)$ associated to operator~(\ref{eqn:2}) is given as:

\smallskip

\begin{equation}\label{eqn:4}
p^{\rm per}(\beta,\mu, \nu, \Delta) = g(z_0) + p^{{\rm id}^\prime}(\beta,\mu) ,
\end{equation}

\bigskip

\noindent
 where $g: \mathbb{R}^+\cup \{ 0\} \rightarrow \mathbb{R} $ is the function defined as:

\[ g(z) =  -a z^2 + (\mu-\Delta) z + 2\nu \sqrt{z}, \]  

\bigskip

\noindent
and $ z_0 (\mu,\nu, \Delta) $ is the unique solution of the equation

\begin{equation}\label{eqn:5}  z + \frac{\Delta -\mu}{2a} = \frac{\nu}{2a\sqrt{z}} \end{equation} for $ z >0.$   

\end{theorem}

\bigskip

\begin{proof} The demonstration of this theorem  is a simple variation of an approach  based on the use of the canonical ensemble, already applied by other authors with similar purposes  (see for example \cite{C2} and refe\-rences therein).

Let $f^{\rm per}_{\ell}(\beta,\varrho),\;f^{\rm id}_{\ell} (\beta,\varrho)$ be the free canonical energies at finite
vo\-lume $V_{\ell}$, inverse temperature $\beta$ and density $\varrho_{\ell},$
corresponding to the perturbed system and the free Bose gas, respectively. 
 
 Let $\;f^{\rm id^{'}}_{\ell} (\beta,\varrho-\varrho_{\mathbf{0}})$ be the finite free canonical
energy asso\-ciated to $ \hat{H}^{0^\prime}.$ 

Being $n_{\mathbf{p}}, N  \in \mathbb{N} \cup \{ 0\}$ for all $\mathbf{p}\in \Lambda^*_{\ell},$  the finite canonical free energies $
f^{\rm id}_{\ell} (\beta,\varrho ), f^{\rm per}_{\ell} (\beta,\varrho )$ can be
written in the following form,

\begin{eqnarray*}
f^{\rm id}_{\ell} (\beta,\varrho) = -\frac{1}{\beta V_{\ell}} \log
\sum_{n_{\mathbf{p}} =0,1,2,\dots,\,\mathbf{p}\in
\Lambda^*_{\ell}}{}&\exp\left(-\beta
\displaystyle\sum_{\mathbf{p}\in\Lambda^*_{\ell}}\lambda_{\ell}(\mathbf{p})
n_{\mathbf{p}}\right) \delta_{\sum_{\mathbf{p}\in
\Lambda^*_{\ell}}n_{\mathbf{p}} =[\varrho V_{\ell}]},
\end{eqnarray*}

\begin{equation*}f^{\rm per}_{\ell}(\beta,\varrho ) = -\frac{1}{\beta V_{\ell}} \log
\left( \displaystyle\sum_{\dots+n_{\mathbf{p}} +\dots=[\varrho
V_{\ell}]} e^{-\beta V_{\ell} h_{\ell}
(\varrho,\varrho_{\mathbf{0}},\varrho_{\mathbf{0}})}\right),\end{equation*}
where
\begin{equation*}\begin{split}
 h_{\ell} (\varrho, \varrho_{\mathbf{0}})
 = \Delta \varrho_{\mathbf{0}} + \varrho^2_{\mathbf{0}} - 2\nu\sqrt{\varrho_{\mathbf{0}}} +\end{split}\end{equation*}

\begin{equation*}\begin{split}
-\frac{1}{\beta V_{\ell}} \log {\sum_{n_{\mathbf{p}}
= 0,1,2,\dots, \mathbf{p}\in \Lambda^*_l\backslash
\{\mathbf{0}\}}}{}&\exp\left(-\beta
\left(\sum_{\mathbf{p}\in \Lambda^*_{\ell}\backslash
\mathbf{0}}
\lambda_{\ell}(\mathbf{p}) n_{\mathbf{p}} \right)\right) \delta_{N^{\prime} = [\varrho V_{\ell}]-[\varrho_0 V_{\ell}]},\end{split}
\end{equation*}
and $ N^{\prime} = \displaystyle\sum_{\mathbf{p}\in
\Lambda^*_{\ell}\backslash \{\mathbf{0}\}}n_{\mathbf{p}}$.
The following inequality

 \begin{equation*}\begin{split} - f^{\rm per}_{\ell} (\beta,\varrho )= &
 \frac{1}{\beta V_{\ell}}\log \left(\displaystyle
 \sum_{..+ n_{\mathbf{p}}+..= [\varrho V_{\ell}] } e^{-\beta V_{\ell} h_{\ell} (\varrho,
 \varrho_{\mathbf{0}})}\right)
 \geq \\& \frac{1}{\beta V_{\ell}} \log \left(e^{-\beta V_{\ell} h_{\ell} (\varrho, \varrho_{\mathbf{0}})}
 \right)=  -h_{\ell}(\varrho, \varrho_{\mathbf{0}}) ,
 \end{split}\end{equation*} holds for $\varrho_{\mathbf{0}} \in [0,\varrho]$,
 being $[b]$ the integer part of $b$. This inequality implies that,
 \begin{equation*}
f^{\rm per} _{\ell}(\beta,\varrho )\leq \displaystyle\inf_{\varrho_{0}
 \in [0,\varrho]} h_{\ell} (\varrho, \varrho_{\mathbf{0}} ).
 \end{equation*}
  On the other hand we have,
\begin{equation*}\begin{split}  - f^{\rm per}_{\ell} (\beta,\varrho ) &
\leq   \frac{1}{\beta V_{\ell}}\log \left(\displaystyle\sum^{[\varrho
V_{\ell}]}_{n_{\mathbf{0}}= 0,N^{\prime}=0} e^{-\beta V_{\ell}
\displaystyle\inf_{\varrho_{\mathbf{0}} \in
[0,\varrho]}h_{\ell}(\varrho, \varrho_{\mathbf{0}})}\right)
\\& \leq \frac{1}{\beta V_{\ell}}\log \left( e^{-\beta V_{\ell}
\displaystyle\inf_{\varrho_{\mathbf{0}} \in
[0,\varrho]} h_{\ell} (\varrho,
\varrho_{\mathbf{0}} )}
\displaystyle\sum^{[\varrho V_{\ell}]}_{n_{\mathbf{0}}=0,N^{\prime}=0
}1\right)
\\& \leq - \inf_{\varrho_{\mathbf{0}}
\in [0,\varrho]} h_{\ell}(\varrho, \varrho_{\mathbf{0}})
+\frac{2}{\beta V_{\ell}}\log \left( 1 +
\frac{[\varrho V_{\ell}](1+ [\varrho V_{\ell}] )}{2}\right)\\& \leq -
\inf_{\varrho_{\mathbf{0}}  \in
[0,\varrho]} h_{\ell} (\varrho, \varrho_{\mathbf{0}}) + \frac{4}{\beta V_{\ell}} \ln ([\varrho
V_{\ell}]+1).\end{split}\end{equation*}
Thus, we obtain the inequalities,

\begin{equation*}\begin{split}
\displaystyle\inf_{\varrho_{\mathbf{0}}
 \in [0,\varrho]}h_{\ell} (\varrho,
\varrho_{\mathbf{0}} )-\frac{4}{\beta
V_{\ell}}\ln( [\varrho V_{\ell}] + 1) \leq f^{\rm per}_{\ell}(\beta,\varrho ) &\\
\leq
\displaystyle\inf_{\varrho_{\mathbf{0}} \in
[0,\varrho]}h_{\ell} (\varrho, \varrho_{\mathbf{0}}). \end{split}\end{equation*}

 Therefore, in the thermodynamic limit it follows that,

\begin{equation*}f^{\rm per}(\beta,\varrho ) = \displaystyle\lim_{V_{\ell}\to\infty}
\inf_{\varrho_{\mathbf{0}} \in
[0,\varrho]}h_{\ell} (\varrho,
\varrho_{\mathbf{0}}).\end{equation*}

 In other words,

\bigskip

\begin{eqnarray*}
f^{\rm per} (\beta,\varrho )&=&
\inf_{\varrho_{\mathbf{0}}\in [0,\varrho]}
\{ \Delta \varrho_{\mathbf{0}} +
a\varrho^2_{\mathbf{0}}  - 2\nu  \sqrt{\varrho_{\mathbf{0}}} + f^{\rm id^{'}}(\beta,\varrho-\varrho_{\mathbf{0}})\},
\end{eqnarray*} where

$$
f^{\rm per} (\beta,\varrho ) = \displaystyle\lim_{V_{\ell},N\to\infty} f^{\rm per}_{\ell}(\beta,\varrho),\;\; f^{\rm id^{'}}(\beta,\varrho-\varrho_{\mathbf{0}}) = \displaystyle\lim_{V_{\ell},N\to\infty} f^{\rm id^{'}}_{\ell} (\beta,\varrho-\varrho_{\mathbf{0}})
$$

\bigskip

Since $f^{\rm per}(\beta,\varrho )$ is a convex function of $\varrho$, its
Legendre transform coincides with the grand canonical limit pressure
$ p^{\rm per}(\beta,\mu, \nu, \Delta)  $, i.e.,
\begin{equation*} p^{\rm per}(\beta,\mu, \nu, \Delta)    =
\displaystyle\sup_{\varrho \geq 0}\{\mu\varrho - f^{\rm per}(\beta,\varrho )\}.
\end{equation*} Hence, for $ \mu \leq 0,\; \beta > 0,$ 
 \begin{eqnarray*}p^{\rm per}(\beta,\mu, \nu, \Delta)   & =
\displaystyle\sup_{\varrho_{\mathbf{0}}\in
[0,\infty)}\{( \mu - \Delta )\varrho_{\mathbf{0}}  -
  a\varrho^2_{\mathbf{0}} + 2\nu  \sqrt{\varrho_{\mathbf{0}}} \} + p^{\rm id^{'}}(\beta,\mu).
\end{eqnarray*}

\bigskip

The continuous function $g: [0,\infty) \to \mathbb{R}$ defined as $$g(\varrho_{\mathbf{0}}) =  - a\varrho^2_{\mathbf{0}} +  ( \mu - \Delta )\varrho_{\mathbf{0}} +2\nu  \sqrt{\varrho_{\mathbf{0}}}, $$ satisfies  $g(0) = 0 $ and $\displaystyle\lim_{\varrho_{\mathbf{0}} \to \pm \infty} g(\varrho_{\mathbf{0}}) = -\infty.$ From this it follows that $g$ has a global maximum for $\varrho_{\mathbf{0}}\geq 0.$ 

Taking $z =  \varrho_{\mathbf{0}}$  we get 
\[
p^{\rm per}(\beta,\mu, \nu, \Delta) =  \displaystyle\sup_{z\geq 0} \{ f(z) \} + p^{ \rm id^{\prime}}(\beta,\mu).
\] 

\bigskip

Therefore, under the hypotheses of the theorem  $f(z) $ attains its ma\-ximum at $ z_0 (\mu), $   positive root  of the equation: 

\[ g^{\prime} (z) =   -2az + (\mu-\Delta) + \frac{\nu}{\sqrt{z}} = 0,\; z >0.\]

On the other hand, $ p^{ \rm id^{\prime}}(\beta,\mu)$ is finite for $\mu\leq 0.$ Thus,

\[
p^{\rm per}(\beta,\mu, \nu, \Delta)  = g(z_0) + p^{ \rm id^{\prime}}(\beta,\mu).
\]

\end{proof}

\bigskip

\begin{corollary}\label{thm:2}   Under the hypotheses of Theorem~\ref{thm:1} the  Bose system with energy ope\-rator $ \hat{H}^{\rm per}_{\ell}$ undergoes  non-conventional  macroscopic occu\-pation of the ground state, being

\begin{equation}\label{eqn:6}
{\rho}^{\rm per}_\mathbf{0} (\mu)= 
z^2_0 (\mu,\nu, \Delta) \end{equation}  

\bigskip

\noindent
the density of particles in the condensate.
\end{corollary}

\begin{proof}
Convexity arguments yield, in the thermodynamic limit,  to the follo\-wing inequalities:  

\bigskip

\begin{equation*}\begin{split}\displaystyle\lim_{V_{\ell}\to\infty} \left<   {\hat{\rho}}^2_{\mathbf{0},\ell} \right>_{\hat{H}_{\ell}(\mu)}=  -\displaystyle\lim_{V_{\ell}\to\infty} \partial_{a}  p^{{\rm per}}_{
\ell}(\beta,\mu)   = \partial_{a}  p^{{\rm per}} (\beta,\mu)  &\\ =z^4_0 (\mu,\nu, \Delta) \geq \displaystyle\lim_{V_{\ell}\to\infty} \left<   {\hat{\rho}}_{\mathbf{0},\ell} \right>^2_{\hat{H}_{\ell}(\mu)} =  \left({\rho }^{\rm per}_{\mathbf{0}} (\mu) \right)^2 \end{split}\end{equation*} and,

\bigskip

\begin{equation*}\begin{split} \displaystyle\lim_{V_{\ell}\to\infty} \left<   \sqrt{{\hat{\rho}}_{\ell, \mathbf{0}}} \right>_{\hat{H}_{\ell}(\mu)}=\frac{1}{2}\displaystyle\lim_{V_{\ell}\to\infty} \partial_{\nu}   p^{{\rm per}}_{
\ell}(\beta,\mu)  = \frac{1}{2}\partial_{\nu}   p^{{\rm per}} (\beta,\mu)  &\\=  z_0 (\mu,\nu, \Delta) \leq \displaystyle\lim_{V_{\ell}\to\infty} \sqrt{\left<   {\hat{\rho}}_{\mathbf{0},\ell} \right>_{\hat{H}^{\rm per} (\mu)}} =  \sqrt{{\rho }^{{\rm per}}_{\mathbf{0}} (\mu)}. \end{split}\end{equation*}

\bigskip

The  bounds for the thermal averages $ \left<   {\hat{\rho}}^2_{\mathbf{0},\ell} \right>_{\hat{H}_{\ell}(\mu)},\;\left<   \sqrt{{\hat{\rho}}_{\ell, \mathbf{0}}} \right>_{\hat{H}_{\ell}(\mu)}, $ fo\-llow from the fact that  $ u(x) = x^2$ is a convex function and  $v(x) = \sqrt{x}$  is a concave function, for $x\geq 0.$

These inequalities lead to  $${\rho}^{\rm per}_\mathbf{0} (\mu)= 
z^2_0 (\mu,\nu, \Delta). $$ 

\bigskip

\end{proof}

\bigskip

As a straightforward consequence of Theorem~\ref{thm:1} and Corollary~\ref{thm:2} it follows that in absence of energy gap, i.e., for $\Delta = 0,\;\mu \leq 0,\;\nu >0,  $ the limit pressure   $p^{\rm per} (\beta,0, \nu,\mu)  $ is given as:

\begin{equation}\label{eqn:7}
 p^{\rm per}(\beta,0, \nu,\mu)  = -az^2_0 + \mu z_0 + 2\nu \sqrt{z_0} + p^{ \rm id^{\prime}}(\beta,\mu),
\end{equation} 

\bigskip

\noindent
where $ z^2_0 (\mu,\nu, 0)$ represents, simultaneously,  the  solution of the equation \begin{equation}\label{eqn:8} -2az  +\mu + \frac{\nu}{\sqrt{z}} = 0,\; z >0, \end{equation} and the fraction of condensed particles.

\bigskip

\section{Remarks}\label{sec:4}

\bigskip

\begin{remarks}\label{thm:3}  $\hat{H}_{\ell}$ and  $\hat{H}^{\rm per}_{\ell}$ are full diagonal self-adjoint operators in momentum representation with respect to the number operators $\hat{n}_{\mathbf{p}},$ therefore the $U(1)$ transformations  $$ \hat{a}^{\dag}_{\mathbf{p}} \to e^{i\phi}\hat{a}^{\dag}_{\mathbf{p}},\;\;\hat{a}_{\mathbf{p}} \to e^{-i\phi}\hat{a}_{\mathbf{p}}, \;\phi\in\mathbb{R} $$ leave the Bose commutation rules and the Hamiltonians invariant (the number of particles is conserved).  This gauge symmetry can be broken by adding the perturbation
 
\[ -2\nu\sqrt{V_{\ell}} ( \hat{a}^{\dag}_{\mathbf{0}} + \hat{a}_{\mathbf{0}}) \] 

\bigskip

\noindent
to $\hat{H}_{\ell,}$ i.e., by defining a new operator  $\hat{H}^{\rm sbr}_{\ell}$ as

\begin{equation}\label{eqn:9}\hat{H}^{\rm sbr}_{\ell}:= \hat{H}_{\ell} -2\nu\sqrt{V_{\ell}} ( \hat{a}^{\dag}_{\mathbf{0}} + \hat{a}_{\mathbf{0}})\end{equation}

\bigskip

From a result due to A. S\"ut\H o in ref.~\cite{SUT} it follows that:

\begin{equation*}\displaystyle\lim_{V\to\infty}  \left\langle  \frac{\hat{a}^{\dag}_{\mathbf{0}}}{\sqrt{V}} \right\rangle_{\hat{H}^{\rm{sbr}}
_{\ell}(\mu)} = \displaystyle\lim_{V\to\infty} \left\langle  \frac{\hat{a}_{\mathbf{0}}}{\sqrt{V}} \right\rangle_{\hat{H}^{\rm{sbr}}
_{\ell}(\mu)} = \nu\sqrt{\rho^{\rm{sbr}}
_{\mathbf{0}}(\mu)} \end{equation*} 

On the other hand,  

\begin{equation*}\displaystyle\lim_{V\to\infty}  \left\langle  \frac{\hat{a}^{\dag}_{\mathbf{0}}}{\sqrt{V}} \right\rangle_{\hat{H}^{\rm{per}}
_{\ell}(\mu)} = \displaystyle\lim_{V\to\infty} \left\langle  \frac{\hat{a}_{\mathbf{0}}}{\sqrt{V}} \right\rangle_{\hat{H}^{\rm{per}}
_{\ell}(\mu)} = 0 \end{equation*} since  $\hat{H}^{\rm per}_{\ell}$ is $U(1)-$ invariant.

\bigskip

Thus, from the above identities and Theorem 4.3 in ref.~\cite{CT} it follows that for fixed para\-meters $\mu \leq 0,$ $\nu >0,$  the  limit pressures of the systems whose energy operators are given by (\ref{eqn:1}) and (\ref{eqn:9}) coincide, i.e.,

\bigskip

\begin{equation}\label{eqn:10}  p^{\rm{per}}(\beta,\mu, \nu, \Delta)  = p^{\rm{sbr}}(\beta,\mu, \nu, \Delta) . \end{equation} 

\bigskip

\end{remarks} 

\bigskip

\begin{remarks}\label{thm:4}  Theorem~\ref{thm:1} can be easily extended  to the case of energy operators of the type:  
\begin{equation}\label{eqn:11}
\bar{H}_{\ell} =  \hat{H}^0_{\ell}+\frac{a}{V_{\ell}}
\displaystyle\sum_{\mathbf{p}\in\Lambda^*_{\ell}} \hat{a}^{\dag 2}_{\mathbf{p}}%
\hat{a}^2_{\mathbf{p}} - 2\nu\sqrt{V_{\ell}\hat{n}_{\mathbf{0}}},
\end{equation} by using the same technique applied in its demonstration. However, in this case,  unlike Theorem~\ref{thm:1},  it is necessary   to prove additionally that  
\begin{equation}\label{eqn:13}\displaystyle\lim_{V_{\ell}\to\infty}\displaystyle\sum_{\mathbf{p}\in
\Lambda^*_{\ell}\backslash \{\mathbf{0}\}}\left\langle
\frac{\hat{n}^2_{\mathbf{p}}}{V_{\ell}^2}\right\rangle_{\hat{H}^{\rm id^{\prime}}_{\ell}
(\mu (\varrho-\varrho_{\mathbf{0}} ))} =0. \end{equation} 
which is a straightforward consequence of the definition of the  so-called  Kac measure  of the ideal Bose gas at finite volume and the fact that for $\mathbf{p},V_{\ell}$ fixed and $r\geq 1,\;r\in \mathbb
\mathbb{Z}^+,$ in the canonical ensemble, the  moments
\begin{equation}\label{eqn:12}
\left\langle
\hat{n}^r_{\mathbf{p}}\right\rangle_{\hat{H}^{\rm id^{\prime}}_l
(\varrho-\varrho_{\mathbf{0}} )}
\end{equation}
are monotonic increasing functions of $\varrho$  (see refs.~\cite{pz,bup}). 

\end{remarks} 

\bigskip

\begin{remarks}\label{thm:5} All the results related to  NCBEC  obtained in ref.~\cite{CT} for a system whose energy operator is the sum of the free gas Hamiltonian  plus the perturbation $-  2\nu\sqrt{V}\sqrt{ \hat{n}_{\mathbf{0}} + 1  },\;\nu > 0,$  can be recovered by a\-pplying the technique used for proving Theorem~\ref{thm:1}.  In such a case, for strictly negative values of $\mu$ and $\nu > 0,$ the density of particles in the condensate is $ \frac{\nu^2}{\mu^2}.$
\end{remarks} 

\bigskip

\section{Final comments}\label{sec:5}

\bigskip

In section~\ref{sec:3}, it has been demonstrated that the inclusion of the disturbance $- 2\nu\sqrt{V_{\ell}\hat{n}_{\mathbf{0}}}$ in the Hamitonian given by~(\ref{eqn:1}) -corresponding to a well-known particle system displaying NCBEC- affects the  density of condensed particles. Moreover, in absence of the  energy shift  induced  by the introduction of the parameter $\Delta$ in the single particle spectrum,  it has been also proven that the studied system undergoes NCBEC. 

These results define a quite different scenario from those represented by energy gaps. Furthermore, it has to be stressed that the limit pressures associated to the models represented by Hamiltonians~(\ref{eqn:2}) and~(\ref{eqn:9})  coincide with each other despite the fact that one of them preserves the $U(1)$ symmetry while the other one does not. In some sense, each system mimics the thermodynamic behavior of its counterpart.
 
However,  although this new context looks promi\-sing, it is not less true that the introduction of this type of disturbances, from the point of view of physics, must have a reaso\-nable justification,  an issue that is far from the purpose of this work. On the other hand, independent on physical consistency and even though these results could seem at first glance counterintuitive, the fact that the systems whose energy operators are of the kind given  by~(\ref{eqn:2}) admit the emergence of NCBEC,  beyond the traditional approach  (energy gaps),  makes this matter per se  mathe\-matically  interesting.  In this context, it is nece\-ssary to point out that possibly  even small disorders (of a random type, for example) could produce similar effects  to those caused by the introduced disturbances.

{\small
\begin{figure}[h]
%Centramos la figura
\begin{center}
\vspace{2cm}
% Incluimos el grafico y definimos el ancho y alto, la
% figura se encuentra en el fichero imagen.jpg
%\includegraphics[height=4cm,width=5cm]{de2.jpg}
 \begin{flushright}
  \textcolor{white}{\Large \textsc{\bf  Digital Editions}}\\
 \textcolor{black}{\Large \textsc{\bf  Departamento de Matem\'aticas}}\\
 \textcolor{black}{\Large \textsc{\bf  Universidad de La Serena (ULS)}}\\
\textcolor{black}{\small \textsc{\bf  Cisternas 1200, La Serena, Chile}}\\
\textcolor{black}{\small \textsc{\bf  edicionesdmatuls@userena.cl}}\\
             \textcolor{black}{\bf\small \url{http://www.dmatuls.cl}}\\[1\baselineskip]
\end{flushright}
%imagen.jpg es el nombre de la imagen que va a aparecer
\end{center}
\end{figure}
}

\begin{minipage}[b]{1\linewidth}
\vspace{1.5cm}

{\small The contents in this document are protected by the Chilean Copyright Law 17.336 -in its current version- Law 28.933, and by international copyright laws. All rights are reserved. Reproduction is authorized for academic and/or educational purposes only. The commerciali\-zation of this draft is not allowed.}\\
\\
\\
\\
\\
%{\small  ISBN XXXXXX}
\\
%\vspace{0.5cm}
%{\textcolor{black}{\small @copyright 2018  Departamento de Matem\'aticas ULS. Chile}}
\end{minipage} 

\end{document}